\documentclass[11pt,preprint]{aastex}
\usepackage{graphicx}

\begin{document}

\title{The first detection of the 232~GHz vibrationally excited H$_{2}$O maser in Orion~KL with ALMA}
\author{Tomoya HIROTA\altaffilmark{1,2}, 
Mi Kyoung KIM\altaffilmark{1}, 
Mareki HONMA\altaffilmark{1,2}} 
\email{tomoya.hirota@nao.ac.jp}
\altaffiltext{1}{National Astronomical Observatory of Japan, Mitaka, Tokyo 181-8588, Japan}
\altaffiltext{2}{Department of Astronomical Sciences, The Graduate University for Advanced Studies (SOKENDAI), Mitaka, Tokyo 181-8588, Japan}

\begin{abstract}
We investigated the ALMA science verification data of Orion~KL and found a spectral signature of the vibrationally excited H$_{2}$O maser line at 232.68670 GHz ($\nu_{2}$=1, 5$_{5,0}$$-$6$_{4,3}$). 
This line has been detected in circumstellar envelopes of late-type stars so far but not in young stellar objects including Orion~KL. 
Thus, this is the first detection of the 232~GHz vibrationally excited H$_{2}$O maser in star-forming regions. 
The distribution of the 232~GHz maser is concentrated at the position of the radio Source~I, which is remarkably different from other molecular lines. 
The spectrum shows a double-peak structure at the peak velocities of $-$2.1 and 13.3~km~s$^{-1}$. 
It appears to be consistent with the 22~GHz H$_{2}$O masers and 43~GHz SiO masers observed around Source~I. 
Thus, the 232~GHz H$_{2}$O maser around Source~I would be excited by the internal heating by an embedded protostar, being associated with either the root of the outflows/jets or the circumstellar disk around Source~I, as traced by the 22~GHz H$_{2}$O masers or 43~GHz SiO masers, respectively. 
\end{abstract}


\section{Introduction}

Water is one of the most abundant interstellar molecules after H$_{2}$ and hence, it is important for interstellar chemistry and physics of molecular clouds \citep[e.g.][]{vandishoeck2011}. 
However, due to the large atmospheric opacity, ground-based observations of the H$_{2}$O lines in radio and infrared bands are almost impossible except for the isotopic species (e.g. HDO and H$_{2}^{18}$O) and strong maser lines. 
In particular, the 6$_{1, 6}$$-$5$_{2, 3}$ transition at 22~GHz (lower state energy, $E_{l}$=642~K) is known to show extremely strong maser emission in circumstellar envelopes (CSEs) around late-type stars, young stellar objects (YSOs) in star-forming regions (SFRs), and active galactic nuclei. 
The 22~GHz maser has been used as a unique probe of dense gas and their dynamics with very long baseline interferometers (VLBI) thanks to its extremely high brightness and compact structure \citep[e.g.][]{chapman2007}. 

Other H$_{2}$O maser lines are also detected in millimeter/submillimeter wavelength \citep{humphreys2007}. 
Lower excitation lines in 183~GHz ($E_{l}$=196~K) and 325~GHz ($E_{l}$=454~K) are detected both in CSEs and around YSOs while some of the higher excitation lines including vibrationally excited lines are detected only in CSEs. 
Multi-transition studies of H$_{2}$O maser lines could be powerful tools to investigate shocked regions in CSEs and YSOs at the highest spatial resolution achieved with VLBI and millimeter/submillimeter interferometers when combined with the theoretical models \citep{neufeld1990, neufeld1991}. 

In this Letter, we report the detection of the vibrationally excited H$_{2}$O maser line at 232.68670~GHz ($E_{l}$=3451~K) in a massive SFR Orion~KL at a distance of 420~pc \citep{hirota2007, kim2008} with the Atacama Large Millimeter/Submillimeter Array (ALMA). 
This line has been detected in late-type stars so far but not in YSOs including Orion~KL \citep{menten1989}. 
Thus, this is the first detection of the 232~GHz H$_{2}$O maser line in YSOs. 

\section{Observation}

We employed the public data obtained with the ALMA science verification (SV) on 2012 January 20. 
They are part of a spectral line survey toward the Orion~KL region at band~6 (215-245~GHz). 
The tracking center position of Orion~KL was set to be R. A. =05h35m14.35s and decl.=$-$05$^{\circ}$22\arcmin35\arcsec.0 (J2000). 
The data consist of several spectral settings and net on-source time for each setting was about 20~minutes. 
The baseline lengths ranged from 17 to 265~k$\lambda$ (from 22 to 345~m) consisted of 16$\times$12~m antennas. 
The primary beam size of each 12~m antenna is about 30\arcsec \ at band 6. 

The spectral resolution of the ALMA correlator was 488~kHz, corresponding to the velocity resolution of 0.60$-$0.65~km~s$^{-1}$ at the observed frequency range. 
Dual polarization data were observed simultaneously. 
We made synthesis imaging with the calibrated data for selected observing frequency ranges by using the Common Astronomy Software Applications (CASA) package. 
The natural weighted beam size was 1\arcsec.7$\times$1\arcsec.4 with the position angle of 171$^{\circ}$. 
The resultant typical rms (root-mean-square) noise level is 0.01$-$0.03~Jy~beam$^{-1}$ for each channel map. 
For comparison, we also made synthesized images of selected lines of methyl formate (HCOOCH$_{3}$) as discussed later. The mapped lines are summarized in Table \ref{tab-lines}. 

In addition to the SV data, we analyzed ALMA cycle 0 data for the continuum emission at band~6 in the Orion~KL region. 
The observation was done in the extended configuration on 2012 April 08 with 17$\times$12~m antennas. 
The observed frequency ranges were 240$-$244~GHz and 256$-$260~GHz. 
The ALMA correlator was set for low resolution wideband continuum observations and the spectral resolution was 15.625~MHz. 
The line emissions were subtracted from the visibility data and effective bandwidth was almost half of the observed frequency range. 
The synthesis imaging was done with the CASA software package. 
The uniform-weighted synthesized beam size was 0\arcsec.74$\times$0\arcsec.56 at a position angle of 101$^{\circ}$. 
The on-source integration time was 30~s and the resultant rms noise level of the continuum image was 7~mJy~beam$^{-1}$. 
Further details will be published in the forthcoming paper (T. Hirota et al., in preparation). 

\section{Results}

First, we inspected the observed spectra of the ALMA SV data around the frequency range close to the 232~GHz H$_{2}$O line. 
As a result, we found a significant spectral feature corresponding to a line-of-sight velocity with respect to the local standard of rest (LSR) of about 11~km~s$^{-1}$. 
We checked the database of molecular lines, Splatalogue\footnote{http://www.splatalogue.net/}, which is a compilation of the JPL, CDMS, and Lovas/NIST catalogs \citep{pickett1998, muller2005, lovas2004}, to confirm possible contamination of other spectral lines. 
We found that the torsionally excited HCOOCH$_{3}$ line at 232.68393~GHz ($v_{t}$=1, 19$_{10,10}$$-$18$_{10,9}$ E) could be another candidate of this line if the source LSR velocity is about 8~km~s$^{-1}$. 
Thus, one should be cautious in identifying the detected lines. 

It is well known that Orion~KL show an enormous amount of molecular lines described as a line forest \citep{beuther2005}. 
The ALMA data also show a number of spectral lines and a significant fraction of them are unassigned to known molecular lines. 
According to previous line survey and interferometer observations, these molecular lines suggest complex velocity/spatial components in Orion~KL such as the Hot Core, the Compact Ridge, SMA1, and Source~I \citep{blake1986, wright1996, beuther2005, favre2011}. 
All of them show different radial velocity, velocity width and spatial distribution. 
In addition, they are known to show significant chemical differentiation. 
For example, oxygen-bearing organic molecules such as HCOOCH$_{3}$ are known to be distributed mainly in the Compact Ridge at an LSR velocity of 8~km~s$^{-1}$ with a linewidth of 2~km~s$^{-1}$, while nitrogen-bearing species tend to peak at the Hot Core at the LSR velocity of 5~km~s$^{-1}$ and with wider velocity widths \citep{blake1986, wright1996, beuther2005, favre2011}. 
On the other hand, the SiO masers are distributed within 100~AU from the radio Source~I \citep{menten1995, reid2007, kim2008}. 
According to the observation with the Very Large Array (VLA), the 22~GHz H$_{2}$O maser features are distributed in more extended regions, but they are concentrated around Source~I, which are called the "shell masers", and the Compact Ridge \citep{gaume1998}. 
If the detected feature really originates from the H$_{2}$O masers at 232.68670~GHz, it can be distinguished based on the distribution and velocity structure of the emitting regions from other thermal lines. 

\begin{figure}
\begin{center}
\includegraphics[width=18cm]{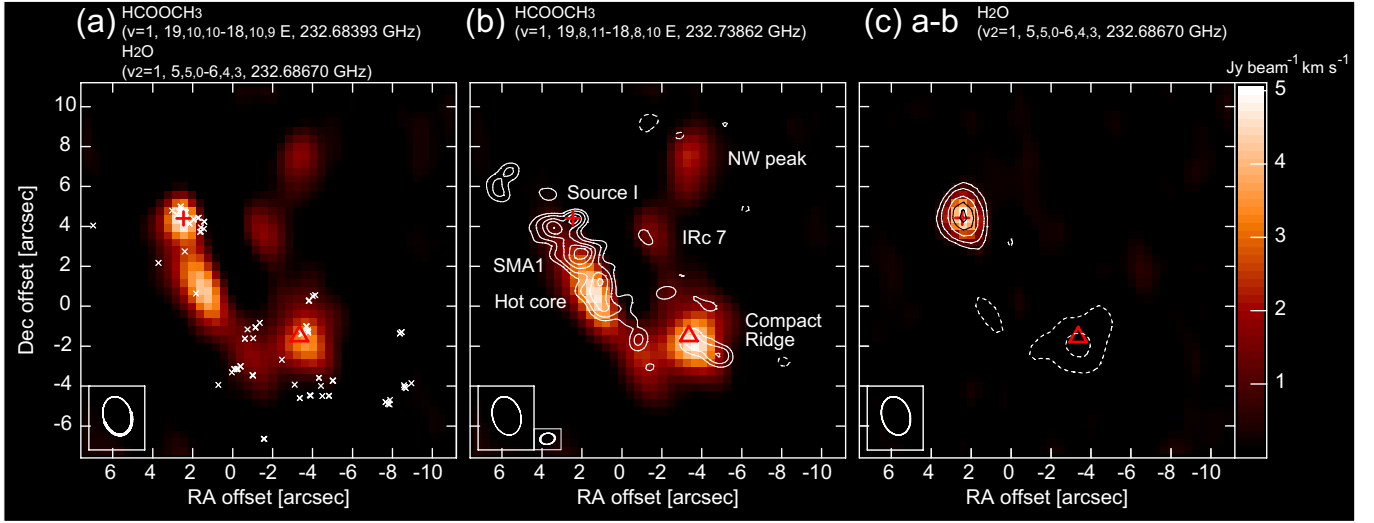}
\caption{
(a) Integrated intensity map of the {\it{blended}} HCOOCH$_{3}$/H$_{2}$O feature with the velocity range of integration of 0$-$12~km~s$^{-1}$. 
The positions of the 22~GHz H$_{2}$O maser spots \citep{gaume1998} are superposed as indicated by white X~marks. 
A red cross and triangle represent the positions of Source~I associated with the SiO and 22~GHz H$_{2}$O masers \citep{gaume1998, kim2008} and the bursting 22~GHz H$_{2}$O maser features \citep{hirota2011} in the Compact Ridge region, respectively. 
(b) Same as panel (a) but for the {\it{pure}} HCOOCH$_{3}$ line. 
The velocity range of integration is 2$-$14~km~s$^{-1}$. 
The dust continuum image (T. Hirota et al., in preparation) is shown as white contours. 
The contour starts at 5$\sigma$ level and its interval of 5$\sigma$, with the 1$\sigma$ noise level of 7~mJy~beam$^{-1}$. 
(c) Difference between the panels (a) and (b). 
The emission component is identified as the 232~GHz H$_{2}$O maser line. 
Contour levels are $-$10, $-$5, 5, 10, 20, 30, and 40 times 0.12~Jy~beam$^{-1}$~km~s$^{-1}$. 
Dotted contours represent negative levels. 
}
\label{fig-map}
\end{center}
\end{figure}

Then we made synthesis images of the spectral feature of the 232.68393~GHz HCOOCH$_{3}$ line and/or the 232.68670~GHz H$_{2}$O line (hereafter we call {\it{blended}} HCOOCH$_{3}$/H$_{2}$O feature) by using the calibrated SV data. 
The results are shown in Figure \ref{fig-map}. 
For comparison, we show a reference image of another HCOOCH$_{3}$ line at 232.73862~GHz ($v_{t}$=1, 19$_{8,11}$$-$18$_{8,10}$ E) having a similar frequency, lower state energy, and expected intensity (hereafter we call this line {\it{pure}} HCOOCH$_{3}$ feature). 
As can be seen in Figure \ref{fig-map}, overall distributions and peak intensities are quite similar. 
For example, both the {\it{blended}} HCOOCH$_{3}$/H$_{2}$O and the {\it{pure}} HCOOCH$_{3}$ maps show four dominant compact condensations coincident with the Hot Core, the Compact Ridge, the Northwest peak, and IRc~7. 
They are consistent with previous HCOOCH$_{3}$ observations \citep{favre2011}. 
One of the brightest peaks is coincident with the Compact Ridge with the integrated intensities of 3.5~Jy~beam$^{-1}$~km~s$^{-1}$ and 4.9~Jy~beam$^{-1}$~km~s$^{-1}$ for the {\it{blended}} HCOOCH$_{3}$/H$_{2}$O feature and the {\it{pure}} HCOOCH$_{3}$ line, respectively. 

However, one can note a striking difference that only the {\it{blended}} HCOOCH$_{3}$/H$_{2}$O feature shows a significant peak at the position of Source~I. 
This source is also associated with the strong SiO masers \citep{reid2007, kim2008} and 22~GHz H$_{2}$O masers \citep{gaume1998}. 
By subtracting the {\it{pure}} HCOOCH$_{3}$ map from the {\it{blended}} HCOOCH$_{3}$/H$_{2}$O map, the residual emission component is clearly concentrated at the Source~I position as shown in Figure \ref{fig-map}(c). 
It is thought to be the contribution from the H$_{2}$O line. A negative component in the Compact Ridge could be due to an intensity variation of the HCOOCH$_{3}$ maps between the two transitions. 
We also imaged five more HCOOCH$_{3}$ lines which are not affected by the contamination from other molecular lines (Table \ref{tab-lines}), and we found that none of them shows a significant peak at the Source~I position. 
Therefore, emission features in the {\it{blended}} HCOOCH$_{3}$/H$_{2}$O map can be distinguished between HCOOCH$_{3}$ and H$_{2}$O lines; HCOOCH$_{3}$ is extended over the Hot Core, the Compact Ridge, the Northwest peak, and IRc~7 while H$_{2}$O is only distributed around Source~I. 

\begin{figure}
\begin{center}
\includegraphics[width=15cm]{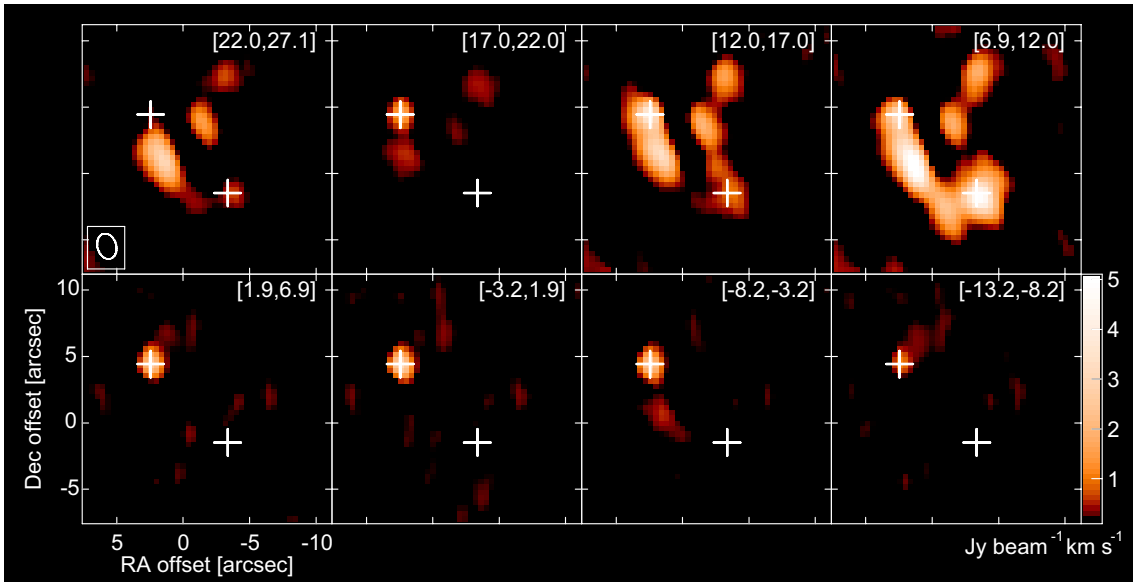}
\caption{Channel maps of the {\it{blended}} HCOOCH$_{3}$/H$_{2}$O feature. 
The data are integrated over eight velocity channels (5~km~s$^{-1}$ widths). 
The velocity range of integration measured with respect to the rest frequency of the 232~GHz H$_{2}$O line, 232.68670~GHz, are labeled at the top-right corner of each panel in unit of km~s$^{-1}$. 
The beam size is indicated in the top-left panel. 
White crosses represent the positions of Source~I and the bursting H$_{2}$O maser feature in the Compact Ridge as shown in Figure \ref{fig-map}. 
Color-scales are set in order to heighten the weaker components outside the Hot Core and the Compact Ridge regions. }
\label{fig-chmap}
\end{center}
\end{figure}

To investigate the spatial and velocity structure of the {\it{blended}} HCOOCH$_{3}$/H$_{2}$O feature more in detail, we made velocity channel maps and the spectra at the selected positions as shown in Figures \ref{fig-chmap} and \ref{fig-spectrum}, respectively. 
The dominant emission components of the {\it{blended}} HCOOCH$_{3}$/H$_{2}$O feature are found in the velocity channels from 6.9 to 17.0~km~s$^{-1}$ as shown in Figure \ref{fig-chmap}. 
They are most likely attributed to the HCOOCH$_{3}$ line blended with the H$_{2}$O line, while higher velocity components are affected by the contamination from the ethyl cyanide ($^{13}$CH$_{3}$CH$_{2}$CN) line at 232.67737~GHz. 
Other weak emission components can be seen in the velocity channels from 17.0 to 27.1~km~s$^{-1}$ and from $-$8.2 to $-$3.2~km~s$^{-1}$. 
They are identified as the methyl acetylene ($^{13}$CH$_{3}$CCH) and acetone ((CH$_{3}$)$_{2}$CO) lines at 232.67074~GHz and 232.69487~GHz, respectively. 
However, the channel maps show notable condensation around Source~I with a velocity range wider than that of the HCOOCH$_{3}$ and other molecular lines. 
These velocity structures can be seen more clearly in the spectra in Figure \ref{fig-spectrum}. 
The HCOOCH$_{3}$ lines toward the Hot Core and the Compact Ridge show narrower linewidths of about 2~km~s$^{-1}$ at the peak velocities of 8~km~s$^{-1}$. 
In contrast, the spectrum of the {\it{blended}} HCOOCH$_{3}$/H$_{2}$O feature shows a double-peaked structure with the velocity range from $-$10 to 20~km~s$^{-1}$. 
Their peak flux densities are derived by the two-component Gaussian fitting to be 0.28$\pm$0.02~Jy and 0.43$\pm$0.02~Jy at the velocity of $-$2.1$\pm$0.6~km~s$^{-1}$ and 13.3$\pm$0.4~km~s$^{-1}$, respectively. 
On the other hand, no spectral feature is detected for {\it{pure}} HCOOCH$_{3}$ toward Source~I. 

Interestingly, the spectral profile of the {\it{blended}} HCOOCH$_{3}$/H$_{2}$O feature toward Source~I appears to be analogous to the 22~GHz shell masers \citep{gaume1998} and the 43~GHz SiO ($v$=1) masers \citep{kim2008} as shown in Figure \ref{fig-maser}. 
One can see a common structure showing double-peaks at almost the same velocities and velocity ranges. 
Since the higher velocity features of the SiO ($v$=1) maser show slightly redshifted with respect to the H$_{2}$O maser lines, the {\it{blended}} HCOOCH$_{3}$/H$_{2}$O feature would have a closer relation with the 22~GHz H$_{2}$O masers. 

It is unlikely that other molecular lines with high velocity components contribute to this peak because no such molecular species is known except the SiO thermal \citep{beuther2005, zapata2012} and maser \citep{reid2007, kim2008} lines, as well as the H$_{2}$O maser lines \citep{gaume1998}. 
Therefore, we can safely conclude that at least the emission feature associated with Source~I is the vibrationally excited H$_{2}$O maser line at 232.68670~GHz. 

\section{Discussion}

\begin{figure}
\begin{center}
\includegraphics[width=8cm]{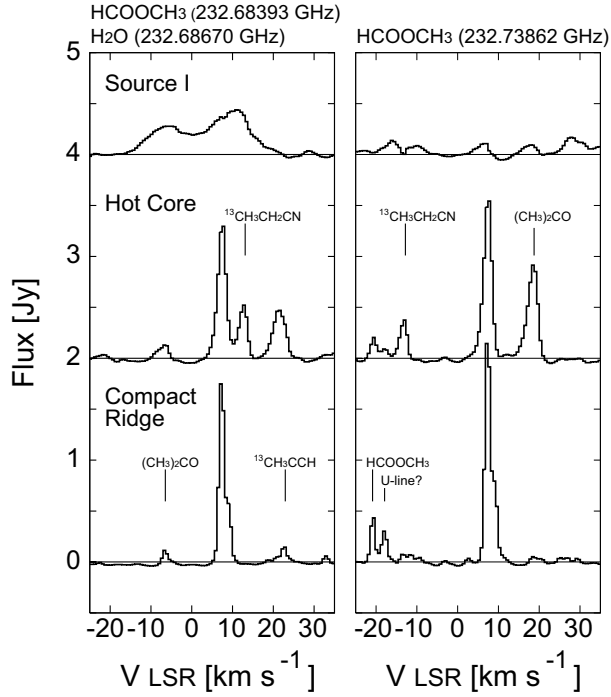}
\caption{Spectra of the {\it{blended}} HCOOCH$_{3}$/H$_{2}$O feature (left) and the {\it{pure}} HCOOCH$_{3}$ line (right) toward the dust continuum peaks at Source~I, the Hot Core, and the Compact Ridge (Figure \ref{fig-map}). 
The flux density is integrated over a 2\arcsec$\times$2\arcsec \ region around each peak. 
For the left panel, the LSR velocities are measured with respect to the rest frequency of the HCOOCH$_{3}$ line, 232.68393~GHz. 
Other lines are identified based on the Splatalogue database as indicated in each panel. }
\label{fig-spectrum}
\end{center}
\end{figure}

\begin{figure}
\begin{center}
\includegraphics[width=8cm]{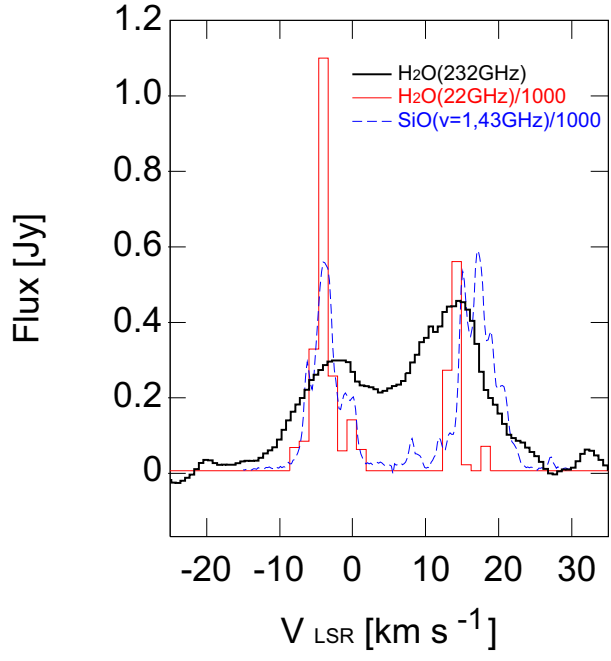}
\caption{Spectra of the 232~GHz H$_{2}$O maser (thick line), 22~GHz H$_{2}$O shell maser \citep[thin line; ][]{gaume1998}, and 43~GHz SiO ($v$=1) maser \citep[dashed line; ][]{kim2008} associated with Source~I. 
These lines are observed with ALMA, VLA, and VERA (VLBI Exploration of Radio Astrometry), respectively. 
Note that the flux scales of the 43~GHz SiO maser and the 22~GHz H$_{2}$O masers are multiplied by a factor of 1/1000. }
\label{fig-maser}
\end{center}
\end{figure}

As discussed above, we can identify the 232.68670~GHz feature detected in the ALMA SV data for Orion~KL as the vibrationally excited H$_{2}$O maser. 
This is the first detection of this maser line in YSOs. 
The peak flux of the {\it{blended}} feature observed with the 2\arcsec$\times$2\arcsec \ aperture is 0.43~Jy (Figures \ref{fig-spectrum} and \ref{fig-maser}). 
It corresponds to the brightness temperature of 2.4~K. 
If this emitting region is as compact as this aperture size, the single-dish observation with the beam size of 30\arcsec \ would yield the brightness temperature of 0.01~K. 
Thus, it was not detectable with the previous observations with single-dish telescopes \citep{menten1989}, although a possible spectral feature can be seen in the line survey data by \citet{sutton1985}, probably attributed to the HCOOCH$_{3}$ line. 

The vibrationally excited H$_{2}$O masers have been detected in only several oxygen-rich late-type stars \citep{menten1989}, which could be attributed to their higher excitation levels (3451~K) than that of the 22~GHz maser (642~K). 
The 232~GHz H$_{2}$O maser around Source~I would be excited due to the internal heating by an embedded YSO as expected from a maser pumping mechanism for late-type stars. 
Source~I is also known as a powering source of the SiO masers, which is quite rare for YSOs \citep{zapata2009}. 
This observational evidence may imply similar characteristics of Source~I and late-type stars. 
Further studies with the millimeter/submillimeter masers in Source~I, along with other YSOs and CSEs in late-type stars, will be crucial in understanding the pumping mechanism of the H$_{2}$O maser lines, physical and dynamical state of these maser sources, and accordingly mass-loss/accretion processes occurring in the YSOs and CSEs. 

The 232.68670~GHz H$_{2}$O maser emission is concentrated around Source~I. 
However, the distribution of the 232~GHz H$_{2}$O maser features could not be resolved with the ALMA SV data with the beam size of 1\arcsec.7$\times$1\arcsec.4. 
According to the double-peaked spectra of the 22~GHz and 232~GHz H$_{2}$O maser as shown in Figure \ref{fig-maser}, the 232~GHz maser features would have similar structure to that of the 22~GHz masers rather than the SiO masers. 
Higher resolution imaging will reveal their spatial structure and provide information about a  possible powering source of the 232~GHz H$_{2}$O maser; 
whether they are really associated with the root of outflows/jets as traced by the 22~GHz H$_{2}$O masers \citep{gaume1998} or with circumstellar disk as traced by the 43~GHz SiO masers \citep{reid2007, kim2008}. 

In the present study, we could not perfectly separate the contribution from the HCOOCH$_{3}$ and the 232~GHz H$_{2}$O maser lines mainly due to the insufficient spatial resolution. 
Therefore, it is still unclear whether the 232~GHz masers are distributed other than in Source~I, such as the Compact Ridge where strong H$_{2}$O maser lines are sometimes detected \citep{hirota2011, gaume1998}. 
A search for the 232.68670~GHz H$_{2}$O maser lines with higher spatial resolution would be important to distinguish their distribution by filtering out the contribution from the thermal and extended HCOOCH$_{3}$ emission. 

Our results clearly demonstrate ALMA's high possibility of detecting new millimeter/submillimeter maser lines with its high sensitivity/resolution. 
Further observational studies with ALMA of millimeter/submillimeter masers will reveal spatial and velocity structure of the maser sources at a resolution of $\sim$10~mas or better, providing a complementary method to the VLBI studies of the SiO and 22~GHz H$_{2}$O masers. 

\acknowledgements
This Letter makes use of the following ALMA data: ADS/JAO.ALMA\#2011.0.00009.SV and 2011.0.00199.S. 
ALMA is a partnership of ESO (representing its member states), NSF (USA) and NINS (Japan), together with NRC (Canada) and NSC and ASIAA (Taiwan), in cooperation with the Republic of Chile. 
The Joint ALMA Observatory is operated by ESO, AUI/NRAO and NAOJ. 
We thank the staff at ALMA for making observations and reducing the science verification data. 
This study is partly supported by the Grant-in-Aid for Young Scientists (A) (24684011). 

{\it Facility:} \facility{ALMA}

{}

\begin{deluxetable}{lcrrl}
\tabletypesize{\scriptsize}
\tablewidth{0pt}
\tablecaption{Molecular Lines Analyzed in the Present Letter
\label{tab-lines}}
\tablehead{
\colhead{Molecule}       & \colhead{Transition}       & \colhead{$\nu$ (GHz)} & \colhead{$E_{l}$ (K)} &
\colhead{Note}
}
\startdata
H$_{2}$O ($\nu_{2}$=1)   & 5$_{5,0}$$-$6$_{4,3}$        & 232.68670             & 3451 & {\it{Blended}} feature  \\
HCOOCH$_{3}$ ($v_{t}$=1)     & 19$_{10,10}$$-$18$_{10,9}$ E & 232.68393	      &  354 & {\it{Blended}} feature  \\
HCOOCH$_{3}$ ($v_{t}$=1)     & 19$_{8,11}$$-$18$_{8,10}$ E  & 232.73862	      &  331 & {\it{Pure}} HCOOCH$_{3}$ \\
HCOOCH$_{3}$ ($v_{t}$=1)     & 18$_{4,14}$$-$17$_{4,13}$ A  & 230.87881             &  290 & Not shown in this Letter  \\
HCOOCH$_{3}$ ($v_{t}$=1)     & 18$_{4,14}$$-$17$_{4,13}$ E  & 231.72416             &  290 & Not shown in this Letter  \\
HCOOCH$_{3}$ ($v_{t}$=1)     & 19$_{10,9}$$-$18$_{10,8}$ E  & 231.74976             &  355 & Not shown in this Letter  \\
HCOOCH$_{3}$ ($v_{t}$=1)     & 19$_{11,9}$$-$18$_{11,8}$ E  & 232.37770             &  368 & Not shown in this Letter  \\
HCOOCH$_{3}$ ($v_{t}$=1)     & 19$_{6,14}$$-$18$_{6,13}$ E  & 235.73208             &  312 & Not shown in this Letter  \\
\enddata
\end{deluxetable}

\end{document}